\batchmode
\documentclass[11pt,english]{article} 
\usepackage[T1]{fontenc}
\usepackage[utf8]{inputenc}
\usepackage{verbatim}
\usepackage{amsmath}
\usepackage{amsthm}
\usepackage{setspace}
\makeatletter

\usepackage{chenpaper}

\makeatother
\newif\ifanon
\anontrue 
\anonfalse 

\newif\ifshowtitle
\newif\ifshowreferences
\newif\ifshowexhibits
\showtitletrue 
\showreferencestrue
\showexhibitstrue

\begin{document}

\ifshowtitle
    \title{\textbf{\color{Black}Most claimed statistical findings in cross-sectional return predictability are likely true}}
    \author{\large {Andrew Y. Chen}\\{\normalsize Federal Reserve Board}}
    \date{November 2025\thanks{First posted to SSRN: August 27, 2021. More recent drafts: \url{https://arxiv.org/abs/2206.15365}. Replication code: \url{https://github.com/chenandrewy/mostly-true}. I thank Alec Erb, Antonio Gil de Rubio Cruz and Rebecca John for excellent research assistance. For helpful comments, I thank Valentin Haddad (the editor), an anonymous referee, Laurent Barras (discussant), Carter Davis (discussant), Wayne Ferson, Dalida Kadyrzhanova, Amit Goyal (discussant), Cam Harvey (discussant), Dino Palazzo, Jeffrey Pontiff (discussant), A. Subra, Russ Wermers, Dacheng Xiu (discussant), David Zeke, Min Zhu and seminar participants at the 2022 ASU Sonoran Winter Finance Conference, Boston College, the Future of Financial Information Webinar, 2024 Midwest Finance Association Meetings, the University of Nebraska, UCLA, UC Riverside, the University of Maryland, 2024 Western Finance Association Meetings, 2024 Chicago Booth Asset Pricing conference, 2024 BIGFI conference, Northeastern University, and Lehigh University. I am grateful to Sterling Yan and Lingling Zheng for sharing their data. The views expressed herein do not necessarily reflect the position of the Board of Governors of the Federal Reserve or the Federal Reserve System.}}

    \ifanon
        \author{}
        \date{}
    \fi

    \maketitle
    \thispagestyle{empty}

    \begin{abstract}
    \begin{singlespace}
    \noindent 
    The false discovery rate (FDR) measures the share of false positives in a set of statistical tests. I develop simple and intuitive bounds on the FDR in cross-sectional predictability publications. The simplest bound requires just a few lines of math and finds $\text{FDR} \le 25\%$ based on summary statistics in eight out of nine previous studies. A more refined bound finds $\text{FDR} \le 9\%$. The FDR is small because randomly selecting accounting ratios produces statistically significant predictability far more often than would occur if there were no predictability. The bounds also reconcile the disparate FDR estimates in the literature.
    \end{singlespace}
    \end{abstract}
    \vspace{1ex}
    \noindent \textbf{\color{Black}JEL Classification}: G0, G1, C1

    \noindent \textbf{\color{Black}Keywords}: stock market predictability,
    stock market anomalies, p-hacking, multiple testing 

    \vspace{10ex}

    \pagebreak{}

\fi

\setcounter{page}{1}

\section{Introduction}

Researchers have discovered hundreds of cross-sectional stock return predictors (\citealt{ChenZimmermann2021}). This abundance has led to concerns about multiple testing. Intuitively, if researchers run many tests, some may be statistically significant by pure chance, even under the null hypothesis of no predictability.

The false discovery rate (FDR) quantifies this problem. The FDR originates from \citet{soric1989statistical}, who defines a ``discovery'' as a statistically significant result, and a ``false discovery'' as a discovery for which the null (unfortunately) holds. Building on this language, \citet{benjamini1995controlling} define the false discovery \emph{rate} as the expected share of discoveries that are false. Discoveries that are not false are true.

This paper provides simple and intuitive bounds on the FDR in the cross-sectional predictability literature. My ``Easy Bound'' is derived in just a few lines of math and can be estimated by plugging in summary statistics from previous studies. This bound shows that the FDR is at most 25\% based on eight out of the nine studies I examine. My ``Visual Bound'' is tighter, and can be calculated by drawing a picture. The Visual Bound implies that the FDR is at most 9\%---in other words, at least 91\% of discoveries are true.

The small FDR comes from a surprising fact: randomly selecting accounting ratios produces statistically significant predictability rather easily. In data mining studies, at least one out of five accounting ratios produces a t-stat exceeding 2.0 (\citealt{yan2017fundamental}; \citealt{chordia2020anomalies}; \citealt{chen2025does}). If there were no predictability, this rate would be one out of twenty---25\% of what is found empirically. My Easy Bound shows that this 25\% is precisely an upper bound on the FDR.

These bounds help reconcile the disparate estimates in the literature. \citet{harvey2016and} and  \citet{chordia2020anomalies} find an FDR exceeding 45\%, a result echoed by \citet{harvey2020false}. In contrast, \citet{chen2018publication} and \citet{jensen2023there} find an FDR close to zero, and \citet{chen2024t} finds a similarly small FDR. My bounds show that these different conclusions are not due to differences in the underlying data. Data in all of these studies imply an FDR of at most 25\%.

Instead, the debate is largely due to differences in interpretation. I explain how \citepos{harvey2016and} own methods imply an FDR of just 9\%. But the paper interprets ``insignificant factors'' as ``false discoveries,'' leading to the claim that most findings are false.  Similar interpretations occur in \citet{harvey2020false}.  \citet{chordia2020anomalies} uses the standard interpretation, but their high FDR estimates likely stem from the omission of key moments from their calibration.

\ifanon
  Replication code is available at [a github url].
\else
  Replication code is available at \url{https://github.com/chenandrewy/mostly-true}.
\fi

\section{An Easy Bound on the FDR\label{sec:ez}}

The cross-sectional predictive power of signal $i$ (a stock-level characteristic) is measured with $\bar{r}_i$. $\bar{r}_i$ is commonly constructed by forming a long-short portfolio based on $i$ and computing a sample mean return, possibly with controls for factor exposure.  The null hypothesis for $i$ is $E(\bar{r}_i)=0$. If this null hypothesis holds, I refer to $i$ as ``null'' and denote this event by $\nullt_i$.  $\bar{r}_i$ has standard error $\SE_i$ and t-stat $t_i \equiv \bar{r}_i/\SE_i$. The samples are long, so the central limit theorem holds:
\begin{align}
    t_i\bigmid\nullt_i \sim \text{Normal}(0,1). \label{eq:ez-null}
\end{align}
Equation \eqref{eq:ez-null} can also be justified using a bootstrap (\citealt{chen2021limits}). These assumptions are standard in cross-sectional asset pricing (\citealt{cochrane2009asset}; \citealt{bali2016empirical}).

In a sense, FDR methods  just add colorful terminology. Following \citet{soric1989statistical}, a ``discovery'' is a significant result. For the bulk of the paper, I use the classical significance threshold of 5\%, which I round to the following t-stat threshold:
\begin{align}
    i \text{ is a discovery if } |t_i|>2. \label{eq:ez-discovery-def}
\end{align}
A ``false discovery'' is a significant signal that is (unfortunately) null.  Significant signals are also called ``predictors'' or ``findings,'' so I also refer to false discoveries as ``false predictors'' or ``false findings.''

With this terminology, a natural definition of the false discovery \emph{rate} is
\begin{align}
    \FDRez & \equiv\Pr\left(\nullt_i\bigmid |t_{i}|>2\right), \label{eq:ez-fdr-def}
\end{align}
where $i$ is randomly selected from some set. In other words, $\FDRez$ is the probability a significant signal is null.  Equation \eqref{eq:ez-fdr-def} is equivalent to the \citet{benjamini1995controlling} definition under weak dependence (e.g. \citealt{chen2024t}). 

Applying Bayes rule and noting $\Pr(\nullt_i) \le 1.0$ leads to an ``Easy Bound'' on the FDR:
\begin{align}
\FDRez & =
    \frac{\Pr\left(|t_{i}|>2\bigmid\nullt_i\right)\Pr\left(\nullt_i\right)}
    {\Pr\left(|t_{i}|>2\right)}
    \label{eq:ez-fdr-bayes}\\
 & \le\frac{\Pr\left(|t_{i}|>2\bigmid\nullt_i\right)}{\Pr\left(|t_{i}|>2\right)}\label{eq:ez-fdr-bound-gen}\\
 &= \frac{5\%}{\Pr\left(|t_{i}|>2\right)},\label{eq:ez-fdr-bound}
\end{align}
where the last line applies Equation \eqref{eq:ez-null}. The Easy Bound is just 5\% divided by the probability a signal is significant.  

It's natural to estimate this probability using method of moments: replace $\Pr\left(|t_{i}|>2\right)$ with the share of signals with $|t_{i}|>2$. The resulting bound is intuitive: if the tail under the null (numerator) cannot explain the actual tail (denominator), then the FDR must be small. 

However, estimating $\Pr\left(|t_{i}|>2\right)$ runs into the problem of publication bias. Since signals with $|t_{i}|>2$ are more likely to be published, the published share of $|t_{i}|>2$ overstates $\Pr\left(|t_{i}|>2\right)$. I address this problem  by considering worst-case scenarios. 

\subsection{Bounding the FDR with Atheoretical Data Mining}\label{sec:ez-dm}

I do not observe the sets of signals researchers consider. I do, however, observe sets of accounting ratios from data mining studies, namely \citet{yan2017fundamental} (YZ); \citet{chordia2020anomalies} (CGS); and \citet{chen2025does} (CLZ). If these accounting ratios are randomly selected, then Equation \eqref{eq:ez-fdr-bound} produces a valid bound on $\FDRez$ from mining accounting ratios. While this is not the ideal bound, it is likely conservative. Given that researchers should use a more refined set of signals, the $\FDRez$ from selecting random accounting ratios should be a worst-case scenario for the cross-sectional literature.

Table \ref{tab:ez-combined}, Panel (a), describes sets of accounting ratios from the YZ, CGS, and CLZ data mining studies. As seen in the ``Accounting Ratios Allowed'' column, these sets are random: they employ little to no economic theory and  instead are designed to systematically explore the data. Nevertheless, at least 20\% of these ratios produce $|t_i| > 2$. Plugging 20\% into Equation \eqref{eq:ez-fdr-bound} implies
\begin{align}
\FDRez & \le\frac{5\%}{0.20}=25\%\label{eq:ez-dm}
\end{align}
for mining accounting ratios. Then, assuming that research does not produce a higher rate of false discoveries, $\FDRez \le 25\%$ for the cross-sectional literature. In other words, at least 75\% of claimed findings are true.

\ifshowexhibits
\begin{table}\caption{\textbf{Easy Bounds on the FDR Based on Nine Studies}}
    \label{tab:ez-combined}
    
    \begin{singlespace}
    \noindent $\FDRez$ is defined in Equation \eqref{eq:ez-fdr-def}.  Papers in Panel (a) begin with $\approx200$ accounting variables and then apply the functional forms in ``Accounting Ratios Allowed.'' Papers in Panel (b) re-examine published predictors. `$\Pr\left(|t_{i}|>2\right)$ Bound' and `Mean Pub t-stat' are inferred from the `Reported Statistic' or from the `Exhibit.' `$\FDRez$ Bound' uses Equation \eqref{eq:ez-dm} (Panel (a)) or Equation \eqref{eq:ez-fdr-extrap} (Panel (b)). `BBH' is \citet{bessembinder2025can}. Using data from nine studies, by eight independent research teams, most claimed findings are true. For eight out of nine studies, $\FDRez \le 25\%$.
    \vspace{1ex}
    
    \end{singlespace}
    \centering{}\setlength{\tabcolsep}{0.5ex} \small
\begin{tabular}{llllcc}
      \toprule
      \multicolumn{6}{c}{Panel (a): Data-Mining as a Worst Case  (Equation \eqref{eq:ez-dm})} \\
      \midrule
            & Data Mining & Accouting & Reported & $\Pr(|t_i|>2)$ & $\FDRez$ \\
            & Study & Ratios Allowed & Statistic & Bound (\%) & Bound (\%) \\
      \midrule
            &       & $X/Y, \: \Delta(X/Y), \: \%\Delta(X/Y),$ & 90th and 10th &       &  \\
      (1)   & Yan and Zheng & $\: \% \Delta X, \: \Delta X / \text{lag} Y, $ & ptile of t-stats & 20    & 25 \\
            & (2017) Table 1 (EW) & $\: \%\Delta X - \%\Delta Y,$  & are 2.41 and -3.48, &       &  \\
            &       & $Y$ restricted to 15 vars & respectively &       &  \\
            &       &       &       &       &  \\
      (2)   & Chordia, Goyal,  & $\Delta X/\text{lag}(X), X/Y$ & 23\% of $|t_i|$ & 23    & 22 \\
            & Saretto (2020) & $(X-Y)/Z$ & exceed 1.96 &       &  \\
            & Table 3 (Alpha) &       &       &       &  \\
            &       &       &       &       &  \\
      (3)   & Chen et al. (2025) & $X/Y, \Delta X/\text{lag}Y$ & Mean t-stat for & 20    & 25 \\
            & Table 1 (Equal- &       & 2nd quintile &       &  \\
            & Weighted) &       & is -2.46 &       &  \\
      \midrule
      \vspace{-1.5ex} &       &       &       &       &  \\
      \midrule
      \multicolumn{6}{c}{Panel (b): Exponential Extrapolation as a Worst Case (Equation \eqref{eq:ez-fdr-extrap})} \\
      \midrule
            & \multirow{2}[2]{*}{Meta Study} & \multirow{2}[2]{*}{Exhibit} & t-stat & Mean Pub & $\FDRez$ \\
            &       &       & Construction & t-stat & Bound (\%) \\
      \midrule
      (4)   & Green, Hand, & Tables 3 & Hand  & 5.1   & 10 \\
            & Zhang (2013) & and 4 & Collected &       &  \\
      (5)   & Chen, Zimm- & Table 2 & Hand  & 4.6   & 11 \\
            & ermann (2020) &       & Collected &       &  \\
      (6)   & Harvey, Liu, & Figure A.1 & Hand  & 4.2   & 12 \\
            & Zhu (2016) & Panel B & Collected &       &  \\
      (7)   & McLean, & Page 16 , & Replicated & 3.6   & 18 \\
            & Pontiff (2016) & Par 1 & In-Sample &       &  \\
      (8)   & Jensen, Kelly,  & Table 1 of & Replicated & 3.6   & 18 \\
            & Pedersen (2021) & BBH 2025 & Full-Sample &       &  \\
      (9)   & Jacobs, & Online App & Replicated & 3.1   & 32 \\
            & Muller (2020) & Table 2 & Full-Sample &       &  \\
      \bottomrule
      \end{tabular}%
      
\end{table}
\fi

This robustness is notable, given that each paper is written by an independent research team. Indeed, each team chose a different set of accounting ratios, as seen in Table \ref{tab:ez-combined}. Nevertheless, they all find that $|t_i| > 2$ for at least 20\% of signals in their main specifications.

While the set of accounting ratios seems to have little effect, a close read of these papers shows that predictability measurement matters. Most importantly, measures that target large stocks (e.g. value-weighted portfolios) produce much smaller t-stats. Section \ref{sec:refined-viz} takes a closer look at this effect, using data on YZ's accounting ratios. Nevertheless, Equation \eqref{eq:ez-dm} provides a conservative estimate of what we'll see there, and can be calculated from easily-available summary statistics.

Studies that focus on randomly-selected past return signals are less common. But the existing evidence also implies a small $\FDRez$. YZ and \citet{chen2024high} examine past return signals as part of their studies, and both find that $|t_i| > 2$ is quite common. Applying Equation \eqref{eq:ez-dm} to YZ's Table 9 shows that the same $\FDRez \le 25\%$ applies to their past return signals. One can also find the same bound using Chen and Dim's publicly-available data. 

\subsection{Bounding the FDR with Conservative Extrapolation\label{sec:ez-extrap}}

A second worst-case scenario comes from conservatively extrapolating the t-stat density into the region affected by publication bias. While extrapolating point estimates is dangerous, we only need to extrapolate an upper bound. That is, we don't need the true density for $|t_i|< 2$---we only need a density that exceeds the true density. As long as this condition is satisfied, the  $\FDRez$ implied by the extrapolation bounds the $\FDRez$ from the research process.

I extrapolate using an exponential distribution for three reasons:
\begin{enumerate}
    \item It fits the observed data well. This is seen in Figure \ref{fig:ez-extrap}, which compares an exponential density with the distribution of t-stats from \citepos{ChenZimmermann2021} replications of 200 published predictors.    

    \item It is conservative. It assumes, in a sense, ``exponentially more'' file drawer t-stats compared to published t-stats. Conservatism is also seen in how the shape is similar to the data-mined  t-stat distribution (dashed line, Figure \ref{fig:ez-extrap}).    

    \item It can be estimated by hand, using summary statistics reported in previous meta-studies. 
\end{enumerate}
An exponential extrapolation is also used in Appendix A.1 of \citet{harvey2016and}.

\ifshowexhibits
\begin{figure}
\caption{\textbf{Conservative Extrapolation from Published t-stats}}
\label{fig:ez-extrap} $t_i$ tests the null that the mean long-short return is zero in a replication of a published predictor from \citet{ChenZimmermann2021}. Solid line extrapolates using an exponential distribution with a mean of 2. Dashed line overlays the distribution of (equal-weighted) absolute t-stats from CLZ's data-mined signals. The extrapolation is conservative, has a similar shape as the data-mined distribution, and fits the data well. 

\vspace{0.15in}
\centering 
\includegraphics[width=0.9\textwidth]{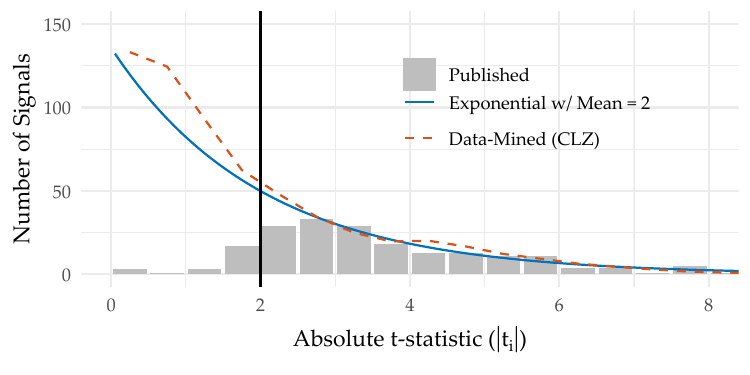} 
\end{figure}
\fi

To understand the third point, note that if $|t_i|$ is exponentially distributed, the memoryless property implies
\begin{align}
E\left(|t_{i}|\right) & =E\left(
    |t_{i}| \bigmid|t_{i}|>2
  \right)
  -2,\label{eq:ez-exp-mean}
\end{align}
that is,  $E(|t_i|)$ can be estimated by subtracting 2 from the mean $|t_i|$ conditional on $|t_i|$ exceeding 2. Since publication generally requires $|t_i|> 2$, the mean published t-stat is not far from this conditional mean.   

This reasoning leads to the following bound:
\begin{align}
\FDRez & \le\frac{5\%}{
    \exp\left(
        -2/(\hat{\mu}_\text{pub}-2)
        \right)
    },
    \label{eq:ez-fdr-extrap}
\end{align}
where $\hat{\mu}_\text{pub}$ is the mean published t-stat and the denominator comes from plugging an exponential distribution with mean $\hat{\mu}_\text{pub}-2$ into Equation \eqref{eq:ez-fdr-bound}.

Panel (b) of Table \ref{tab:ez-combined} applies Equation \eqref{eq:ez-fdr-extrap} to a variety of studies. For example, Chen and Zimmermann's \citeyearpar{chen2020publication} Table 2 reports that the mean hand-collected t-stat across 77 published predictors is 4.6. Equation (\ref{eq:ez-fdr-extrap}) then implies $\FDRez \le 5\% / \exp(-2/(4.6-2)) = 11\%$. Replications and post-publication data tend to result in lower t-stats, leading to higher FDR bounds using \citepos{mclean2016does} and \citepos{jacobs2020anomalies} replications. Nevertheless, the results across six independent teams are consistent. All imply that most findings are true. 

The entry for \citet{harvey2016and} (HLZ) is notable. It shows that the Easy Bound, applied to HLZ's dataset, implies an FDR of at most 12\%. This result contrasts with HLZ's conclusion that most findings are false, and suggests that the disparate FDR estimates in the literature are not due to differences in data. We will see that different methods are not the issue either, and that the debate is driven by different interpretations of multiple testing methods (Section \ref{sec:discuss}).

In summary, the FDR is small because the research process readily generates signals with large t-stats, much more readily than compared to a standard normal distribution. This property can be seen in the relatively high frequency of large t-stats in data mining experiments or in conservative extrapolations of published t-stats. 

\section{More Refined Estimates\label{sec:refined}}

The bounds in Section \ref{sec:ez} are coarse. Rather than bound $\Pr\left(\nullt_i\right)$ based on data, they use the theoretical bound $\Pr\left(\nullt_i\right) \le 1.0$. They gloss over important details of predictability measurement like equal- vs value-weighting. They use only summary statistics from published tables, rather than actual datasets.

This section addresses these issues. Section \ref{sec:refined-viz} bounds $\Pr\left(\nullt_i\right)$ with data. Section \ref{sec:refined-vwffn} conditions the bound on the details of predictability measurement. These methods are applied to the  \citet{chen2025does} and \citet{yan2017fundamental} datasets. Along the way, Section \ref{sec:refined-viz-corr} discusses correlations and standard errors.

\subsection{A Tighter, Visual Bound}\label{sec:refined-viz} 

To obtain a tighter bound, we can use data to make inferences on $\Pr\left(\nullt_i\right)$. A natural data point to use is the fraction of t-stats that are small. Intuitively, if $\Pr\left(\nullt_i\right)$ is large, then many t-stats should be small.

To formalize this intuition, apply the law of total probability to the event that $|t_i|$ is small, say less than 0.5:
\begin{align}
    \Pr\left(|t_i| < 0.5\right)
    &=
    \Pr\left(|t_i| < 0.5\bigmid\nullt_i\right)\Pr\left(\nullt_i\right)
    +
    \Pr\left(|t_i| < 0.5\bigmid\alt_i\right)\Pr\left(\alt_i\right)
    .
    \label{eq:viz-1}
\end{align}
where $\alt_i$ is the event that $i$ is not null. Noting that $\Pr\left(|t_i| < 0.5\bigmid\alt_i\right)\Pr\left(\alt_i\right) \ge 0$ and applying Equation \eqref{eq:ez-null} yields
\begin{align}
    \Pr\left(|t_i| < 0.5\right)
    & \ge \Pr\left(|t_i| < 0.5\bigmid\nullt_i\right)\Pr\left(\nullt_i\right)
    \label{eq:viz-1b}
    \\
    &= (0.38)\Pr\left(\nullt_i\right)
    \label{eq:viz-2}
\end{align}
where 0.38 is the probability that a standard normal r.v. falls between -0.5 and +0.5. Equation \eqref{eq:viz-2} provides a bound on $\Pr\left(\nullt_i\right)$ based on the observed frequency of small t-stats.

Plugging Equation \eqref{eq:viz-2} into Equation \eqref{eq:ez-fdr-bayes} implies a ``Visual Bound'' on $\FDRez$:
\begin{align}
    \FDRez
    \le \left[
        \frac{5\%}{\Pr\left(|t_{i}|>2\right)}
     \right]
     \left[
        \frac{\Pr\left(|t_i| < 0.5\right)}{0.38}
     \right].
     \label{eq:viz-fdr-bound}
\end{align}
This expression tightens the Easy Bound (first bracket) with a data-based bound on $\Pr\left(\nullt_i\right)$ (second bracket). As in the Easy Bound, Equation \eqref{eq:viz-fdr-bound} can be estimated with method of moments: replace probabilities with observed frequencies. This estimation is a variation of \citepos{storey2002direct} Algorithm 1, though my argument does not require independence. I return to this point in Section \ref{sec:refined-viz-corr}.

To obtain numbers for Equation \eqref{eq:viz-fdr-bound}, I use data from \citet{chen2025does} (CLZ), which is publicly available.\footnote{The CLZ data can be found at {https://sites.google.com/site/chenandrewy/}} It's important to use data-mined rather than published signals, since Equation \eqref{eq:viz-fdr-bound} requires reliable data on signals with small t-stats. I use equal-weighted raw returns to match the modal specification in the literature (\citealt{green2013supraview}) but Section \ref{sec:refined-vwffn} examines other performance measures. 9,700 of 29,000 signals have $|t_i| > 2$, and 6,300 have $|t_i| < 0.5$. Plugging these numbers into Equation \eqref{eq:viz-fdr-bound} implies
\begin{align}
    \FDRez
    \le \left[
        \frac{5\%}{9,700/29,000}
     \right]
     \left[
        \frac{6,300/29,000}{0.38}
     \right]
     = 8.5\%
     \label{eq:viz-fdr-num}
\end{align}
for CLZ's data mining process. Then assuming that research does not produce a higher rate of false discoveries, $\FDRez \le 8.5\%$ for the cross-sectional literature. In other words, at least 91.5\% of equal-weighted raw return findings in the literature are true.

I call Equation \eqref{eq:viz-fdr-bound} the Visual Bound because it can be estimated using the following algorithm:
\begin{algorithm}[H]
    \caption{Visual Bound Estimation}
    \label{alg:viz}
    \begin{enumerate}
        \item Plot the data histogram for $|t_i|$ for some data-mined signals.
        \item Draw the largest null histogram that still fits inside the data.
        \item Draw a vertical line at 2.0. To the right of this line, the ratio of the null area to the data area estimates the bound on $\FDRez$ in Equation \eqref{eq:viz-fdr-bound}.
    \end{enumerate}
\end{algorithm}
This algorithm is equivalent to estimating Equation \eqref{eq:viz-fdr-bound} with method of moments, under the assumption that the density of $|t_i|$ is decreasing and that the histogram bins include $[0, 0.5)$. To see this, note that estimating Equation \eqref{eq:viz-fdr-bound} is equivalent to taking the ratio of the right tail areas of two histograms: (1) a null histogram scaled by an estimate of $\Pr\left(|t_i| < 0.5\right) / 0.38$, and (2) a data histogram. Step 2 imposes a scaling factor on the null histogram. This scaling factor is equal to $\Pr\left(|t_i| < 0.5\right) / 0.38$ under the two assumptions. Step 3 then compares the scaled histogram tails.

Figure \ref{fig:viz}, Panel (a), applies Algorithm \ref{alg:viz} to CLZ's equal-weighted raw returns. The gray bars show the data histogram of $|t_i|$. The red bars show the largest null distribution that still fits inside the data. The result is an intuitive visualization of Equation \eqref{eq:viz-fdr-num}. To the right of the vertical line at 2.0, the ratio of the null area (red) to the gray area is only 8.5\%. 

\ifshowexhibits
\begin{figure}
    \caption{\textbf{A Visual Bound on the FDR}} \label{fig:viz}
    Bars show the distribution of CLZ equal-weighted raw return t-stats. The red component is $\text{Normal}(0,1)$ scaled by a constant. Panel (a) uses Algorithm \ref{alg:viz} to determine the constant. Panel (b) uses the Easy Bound. Error bars show one standard error calculated with a cluster bootstrap. Algorithm \ref{alg:viz} provides an visual interpretation of Equation \eqref{eq:viz-fdr-bound} and shows that at least 91\% of equal-weighted raw return findings are true.     
    
    \vspace{0.15in}\centering 
    \subfloat[Visual Bound]{
    \includegraphics[width=0.9\textwidth]{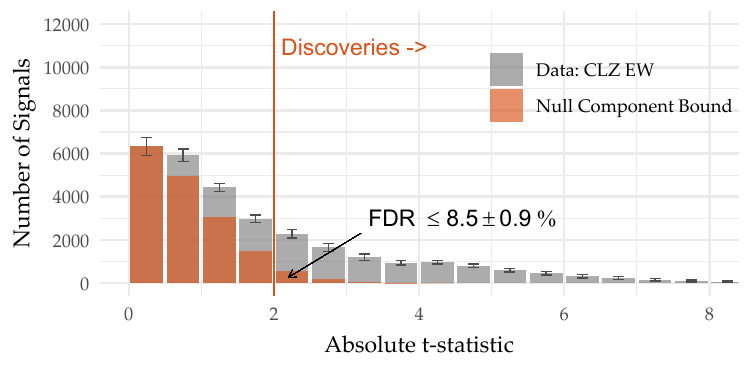}
    } \\    
    \subfloat[Easy Bound]{
    \includegraphics[width=0.9\textwidth]{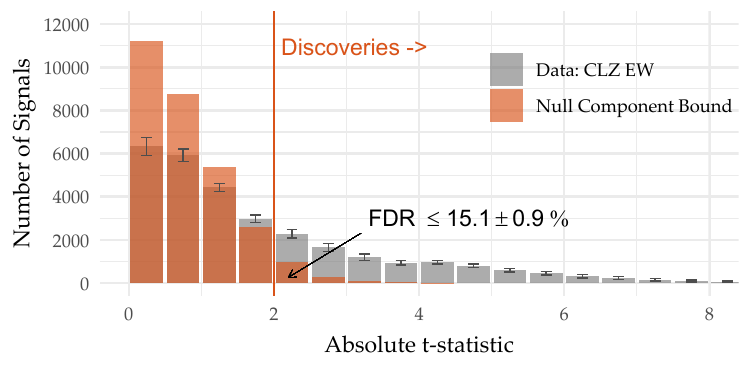}
    } 
\end{figure}
\fi

More broadly, Algorithm \ref{alg:viz} shows how to interpret FDR methods as a histogram decomposition. This interpretation is seen in Figure \ref{fig:viz}, but it can be derived from first principles.

To formalize this interpretation, let $b$ be a histogram bin, and apply the law of total probability to the event that $|t_i|$ falls in bin $b$:
\begin{align}
    \underbrace{\Pr\left(|t_i|\in b\right)
    }_{
        \left[\text{Data Component}\right]_b
    }
    &= \underbrace{
        \Pr\left(|t_i|\in b\bigmid\nullt_i\right) 
        \Pr\left(\nullt_i\right)
    }_{
        \equiv \left[\text{Null Component}\right]_b
    }
    + \underbrace{
        \Pr\left(|t_i|\in b\bigmid \alt_i\right) \Pr\left(\alt_i\right)
    }_{
        \equiv \left[\text{Alternative Component}\right]_b
    }
    .\label{eq:viz-total-prob}
\end{align}
This equation says that each bar of a t-stat histogram (LHS) is the sum of a null component and an alternative component (RHS). We do not know the null component. But Algorithm \ref{alg:viz} can bound it, using an intuitive constraint: the null component cannot be larger than the data component. If it were larger, then the law of total probability would be violated.

The null component, in turn,provides the key quantities in FDR methods. It provides  $\FDRez$:
\begin{align}
    \FDRez 
    &= 
    \frac{
        \sum_{b \hspace{0.2ex}\subseteq[2,\infty)}
        \left[\text{Null Component}\right]_b
    }{
        \sum_{b \hspace{0.2ex} \subseteq[2,\infty)}
        \left[\text{Data}\right]_b
    },\label{eq:viz-fdr}
\end{align}
where I've assumed that the bins have an edge at 2.0. To see why Equation \eqref{eq:viz-fdr} holds, plug in the definition of the null component, and compare to Equation \eqref{eq:ez-fdr-bayes}. The null component also provides $\Pr(\nullt_i)$:
\begin{align}
    \Pr(\nullt_i)
    &=
    \sum_{b \in B} \left[\text{Null Component}\right]_b
    \label{eq:viz-prnull}
\end{align}
where I've assumed that $B$ covers the support of $|t_i|$. Algorithm \ref{alg:viz} provides upper bounds on these null components, and thus upper bounds on $\FDRez$ and $\Pr(\nullt_i)$. 

Equation \eqref{eq:viz-prnull} is visualized in Panel (a) of Figure \ref{fig:viz}: the ratio of the total red area to the total gray area is an upper bound on $\Pr(\nullt_i)$. This ratio implies $\Pr(\nullt_i) \le 56\%$, which may be surprising. $\Pr(\nullt_i) \le 56\%$ means that at least 44\% of randomly-selected accounting ratios have real, non-zero expected returns. 

Though this result may be surprising, a larger $\Pr(\nullt_i)$ is implausible, as illustrated in Panel (b). This panel shows an alternative decomposition, in which $\Pr(\nullt_i) > 56\%$ is possible. The result is an unappealing decomposition, in which red bars exceed the gray bars. This feature is not only visually unappealing, it also violates principles of probability. Yet it is precisely the decomposition implicit in the Easy Bound (Equation \eqref{eq:ez-fdr-bound}). Thus, while the FDR bounds in Section \ref{sec:ez} are easy to calculate, they are likely too conservative. They violate the ``budget constraint'' imposed by the law of total probability.

\subsection{Correlations, Standard Errors, and Factor Structure}\label{sec:refined-viz-corr} 

The Visual and Easy Bounds do \emph{not} require independence. As seen in Algorithm \ref{alg:viz}, FDR methods boil down to histogram estimation, which is valid under the weak dependence assumptions described in \citet{wooldridge1994estimation}. 

Correlations do increase estimation uncertainty. Figure \ref{fig:viz} captures this effect with the \citet{fama2010luck} cluster bootstrap, which preserves cross-sectional correlations. The error bars are non-negligible but the histogram is well-estimated overall. The standard error on the FDR bound is only 0.9 percentage points. In fact, this standard error is likely overstated, as the \citet{fama2010luck} method suffers from ``bootstrap dilation'' in this large scale setting (Efron \citeyear{efron2003second}; \citeyear{efron2010correlated}a; \citeyear{efron2010rejoinder}b).\footnote{%
Correlations also imply that FDR point estimates are too conservative (\citealt{schwartzman2011effect})
} %

A small standard error is consistent with the large number of signals and moderate correlations in the data. Consider Panel (b), which builds on the estimate of $\widehat{\Pr(|t_i|>2)} = 0.33$. Under independence, the standard error would be only $\sqrt{0.33(1-0.33)/29,000} = 0.3$ percentage points, due to the 29,000 signals.\footnote{%
A small standard error is also seen applying the delta method to the Easy Bound, which implies a standard error of $5(1/0.33)^2 \sqrt{0.33(1-0.33)/29,000} \approx 0.1$ percentage points.
} %
Correlations imply a larger standard error, but the correlations in cross-sectional predictors are not particularly high. CLZ find that among data-mined predictor returns, more than 80\% of pairwise correlations are less than 0.25 in absolute value, and that roughly 50 principal components are required to explain 75\% of total variance (see also \citealt{mclean2016does}; \citealt{ChenZimmermann2021}).  

Correlations also impact the interpretation. Figure \ref{fig:viz} shows roughly 10,000 true discoveries, but they cannot all be independent. If they were, one could combine them to form a portfolio with a Sharpe ratio $\sqrt{10,000} = 100$ times larger than the individual Sharpe ratios. In reality, combining data-mined predictors leads to annualized out-of-sample Sharpe ratios of about 1.5 (YZ, CLZ, \citealt{chen2024high}), which, while high, are far short of the $100\times$ boost. The observed Sharpe ratios are broadly consistent with CLZ's principal components analysis, which implies roughly 50 independent discoveries rather than 10,000.

\subsection{Controlling for Value Weighting and Factor Exposure\label{sec:refined-vwffn}}

There are multiple ways to measure the predictive power of a signal $i$. A more refined estimate of $\FDRez$ should control for details of this measurement. Table \ref{tab:yz-fdr} examines how stock weighting and factor model adjustments affect Equation \eqref{eq:viz-fdr-bound}, applied to accounting ratios from the YZ dataset.  Similar results are found using the CLZ data (see the Github replication code).

The first three columns construct $|t_i|$ using equal-weighted long-short decile returns, adjusted for exposure to various factor models.  The factor model adjustments have little impact on the share of $|t_{i}|>2$ or the share of $|t_i| < 0.5$. Thus, the $\FDRez$ bounds are close to the estimates based on raw returns (Figure \ref{fig:viz}), and range from 7.3\% to 9.6\%.

\ifshowexhibits

\begin{table}
\caption{\textbf{FDR Estimates Controlling for Value Weighting and Factor Adjustments}}
\label{tab:yz-fdr}

\begin{singlespace}
\noindent Data consists of long-short decile strategies formed on 18,000 accounting ratios constructed by \citet{yan2017fundamental}. $t_i$ tests the null that alpha = 0 using the stock weighting and the factor adjustment in the column header.  `4-Fac' is FF3+Momentum. Panel (a) uses Equation \eqref{eq:viz-fdr-bound}  to bound $\FDRez$. `$\Pr\left(\nullt_i\right)$ Upper Bound' uses Equation \eqref{eq:viz-2}. Panel (b) uses Equation \eqref{eq:HL-1} to control the FDR following \citet{storey2002direct}, as implemented in \citet{harvey2020false}. After adjusting for value weighting and factor exposure, at least 77\% of findings are true under most specifications, though the bound falls to 59\% controlling for FF3 + Momentum exposure. Harvey and Liu's finding of near-zero ``true strategies'' can be reconciled by interpreting significant factors as ``true factors'' and focusing on the most extreme estimate based on the six specifications in YZ. \vspace{2ex}
\end{singlespace}
\centering{}\setlength{\tabcolsep}{1.2ex} \small

\begin{tabular}{lrrrrrr}
\toprule
& \multicolumn{3}{c}{Equal-Weighted}  &  \multicolumn{3}{c}{Value-Weighted} \\
 & CAPM & FF3 & 4-fac & CAPM & FF3 & 4-fac \\
\cline{2-7}
\vspace{-1.5ex} \\  \multicolumn{7}{l}{Panel (a): Visual Bound on $\FDRez$} \\ \midrule
Share of $|t_i| > 2.0$ & 32.7 & 34.9 & 29.3 & 17.9 & 17.1 & 10.5\\
Share of $|t_i| < 0.5$ & 21.4 & 19.3 & 21.5 & 25.8 & 29.6 & 33.4\\
$\Pr(\nullt_i)$ Upper Bound & 56.4 & 50.8 & 56.5 & 68.0 & 77.9 & 87.8\\
$\FDRez$ Upper Bound & 8.6 & 7.3 & 9.6 & 19.0 & 22.8 & 41.7\\
\\ \multicolumn{7}{l}{Panel (b): Significance using Storey (2002) with $\text{FDR}\le 10\%$} \\ \midrule
$t$-statistic Hurdle $h$ & 1.8 & 1.7 & 1.9 & 2.4 & 2.5 & 3.3\\
Percent Significant & 36.1 & 41.3 & 31.3 & 11.3 & 9.2 & 1.0\\
\bottomrule
\end{tabular}

\end{table}

\fi

The other columns examine value-weighted returns, which do make a difference. While equal-weighted returns consistently imply a share of $|t_i|>2$ of around 30\%, the share under value weighting is much smaller, ranging from 11\% to 18\%. This leads to larger $\FDRez$ bounds ranging from 21\% to 42\%, depending on the choice of factor model. Nevertheless, the Visual Bound still implies that the majority of findings are true, even under value weighting. Notably, findings based on value-weighted CAPM alphas are at least 81\% likely to be true discoveries.

The 42\% FDR bound for the value-weighted 4-factor (FF3 + momentum) alpha is notable. This high FDR  is consistent with weaker predictability found in large and liquid stocks (\citealt{chen2023zeroing}), and the fact that momentum shows up in all size groups (\citealt{fama2008dissecting}). One interpretation is that liquidity and momentum are important for understanding cross-sectional predictability, and potentially account for 42\% of findings in the literature. Nevertheless, liquidity and momentum still leave the majority of predictability unexplained.

\section{Reconciliation with the Literature}\label{sec:discuss} 

The bounds in Sections \ref{sec:ez} provide a partial reconciliation. Table \ref{tab:ez-combined} shows that $\FDRez \le 25\%$, based on many previous studies. This finding is consistent with YZ; \citet{chen2018publication}; \citet{jensen2023there}; and \citet{chen2024t}; who find either a small FDR or many valid predictors.

This section aims to complete the reconciliation. I re-examine  papers that find $\FDRez \gg 25\%$, specifically \citet{harvey2016and}, \citet{harvey2020false}, and \citet{chordia2020anomalies}.  I close by discussing the literature on out-of-sample tests (\citealt{mclean2016does}; YZ; \citealt{linnainmaa2018history}; CLZ). 

Unlike previous sections where significance is defined as $|t_i|>2.0$, this section considers the alternative hurdles. Accordingly, ``FDR'' in this section refers to notions of significance beyond $|t_i|>2.0$. ``$\FDRez$'' still refers to the classical $|t_i|>2.0$.

\subsection{Reconciliation with \citet{harvey2016and} (HLZ)}\label{sec:discuss-hlzfdr}

HLZ's abstract states: ``A new factor needs to clear a much higher hurdle, with a t-statistic greater than 3.0. We argue that most claimed research findings in financial economics are likely false.'' Reconciling this claim with my estimates centers around the interpretation of terms like ``insignificant'' and ``false discovery.''

To understand the interpretations, consider the model used in HLZ's simulated method of moments (SMM) estimates (Section 4.1, page 30). HLZ's model adds parametric assumptions to the framework from Sections \ref{sec:ez} and \ref{sec:refined}. It models $t_i$ using
\begin{align}
    t_i \mid \mu_i &\sim \text{Normal}\!\left( \tfrac{\mu_i}{\sigma/\sqrt{N}},\, 1 \right)
    \label{eq:hlz-1}
    \\
    \Corr\left(t_i, t_j\right) &= \rho \quad \text{for } i \neq j,
    \label{eq:hlz-1b}
\end{align}
where $\mu_i = E(\bar{r}_i)$ is the expected return of factor $i$, $\rho$ is the correlation between factor returns, $\sigma$ is the volatility of factor returns, and $N$ is the length of the sample. $\mu_i$ is modeled using
\begin{align}
    \mu_i &= 0 \qquad\qquad\qquad\quad \text{with prob } p_0 
    \label{eq:hlz-2}
    \\
    \mu_i &\sim \text{Exponential}(\lambda) \quad \text{with prob } (1 - p_0),
    \label{eq:hlz-3}
\end{align}
where $p_0$ is a constant and $\lambda$ is the scale parameter of the exponential distribution. 

In the standard interpretation, a factor is null if $\mu_i=0$, and it is significant if $|t_i|>h$, where $h$ is a significance hurdle. ``Null'' and ``significant'' are related through Equation \eqref{eq:hlz-1}, which implies that $\mu_i$ and $|t_i|$ are correlated. Nevertheless, they are distinct concepts. A null factor may very well have $|t_i|>h$. An alternative factor (drawn from Equation \eqref{eq:hlz-3}) may have $|t_i|<h$. Multiple testing controls add colorful language (a ``discovery'' is a significant factor), but  the distinction between null and significant remains (e.g. \citealt{benjamini1995controlling}).  This distinction is embedded in Principle 2 of the American Statistical Association's ``Statement on Statistical Significance and P-Values'' (\citealt{wasserstein2016asa}). 

In HLZ, this distinction is removed. Section 3.2 ``A multiple testing framework'' states: ``[i]n a factor testing exercise, the typical null hypothesis is that a factor is not significant. Therefore, a factor being insignificant means the null hypothesis is `true' '' (page 13). The same page defines ``false discoveries'' as when ``we conclude a factor is `significant' when it is not,'' and this interpretation is also used in the conclusion (page 37). These definitions imply that HLZ consider a ``false discovery'' a factor that meets the classical 2.0 hurdle, but fails to meet an alternative hurdle selected by HLZ. Using the 3.0 hurdle from HLZ's abstract, factors that satisfy $2.0 < |t_i| < 3.0$ are ``false discoveries''---regardless of the value of $\mu_i$. 

To visualize this distinction, consider Figure \ref{fig:hlz-interp}, which simulates Equations \eqref{eq:hlz-1}-\eqref{eq:hlz-3} using HLZ's SMM estimates (Table 5, $\rho = 0.2$). These estimates come from selecting $\rho=0.2$, $\sigma=0.15/\sqrt{12}$ per month, and $N=240$ to match basic asset pricing facts, and then choosing $p_0=0.444$ and $\lambda=0.555$ to fit the cross-factor distribution of t-stats. I run 1,000 simulations of 10,000 factors. The figure shows 1,378 factors (HLZ's preferred number, Table 5) drawn from across these simulations, each represented by one marker.\footnote{I simulate more factors than HLZ's preferred number to reduce variation in the Monte Carlo calculations. The FDR is invariant to the number of factors.}

\ifshowexhibits
\begin{figure}[h!]\caption{\textbf{Two Interpretations of Harvey, Liu, and Zhu's (2016) SMM Estimates }}
    \label{fig:hlz-interp} I simulate HLZ's SMM estimates (Equations \eqref{eq:hlz-1}-\eqref{eq:hlz-3}). ``Expected Return'' is $\mu_i$, ``t-statistic'' is $t_i$, and each marker represents one factor. ``FDR = 5\%'' and ``FDR = 1\%'' show t-hurdles from HLZ's Table 5. Panel (a) uses the standard interpretation (\citealt{benjamini1995controlling}). Panel (b) follows HLZ, which interprets insignificant factors as ``false discoveries.''  The standard interpretation finds $\FDRez = 9\%$, in-line with Figure \ref{fig:viz}. HLZ's interpretation leads to ``false discoveries''  with expected returns of 12\% per year.
    
    \vspace{0.15in}
    \centering 
    \subfloat[Standard Interpretation]{\includegraphics[width=0.75\textwidth]{%
        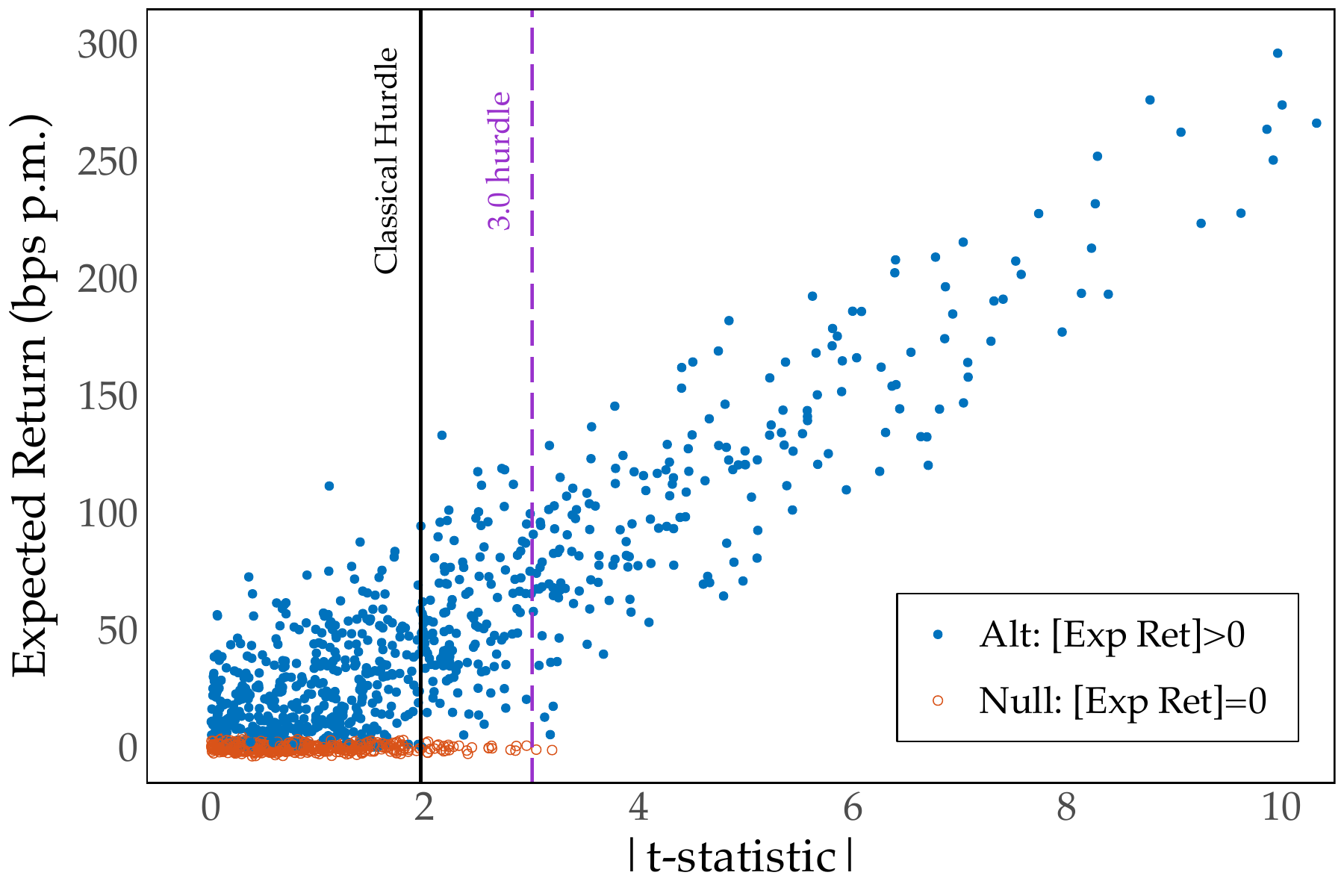
    }} \\
    \subfloat[Harvey, Liu, Zhu's (2016) Interpretation]{\includegraphics[width=0.75\textwidth]{%
        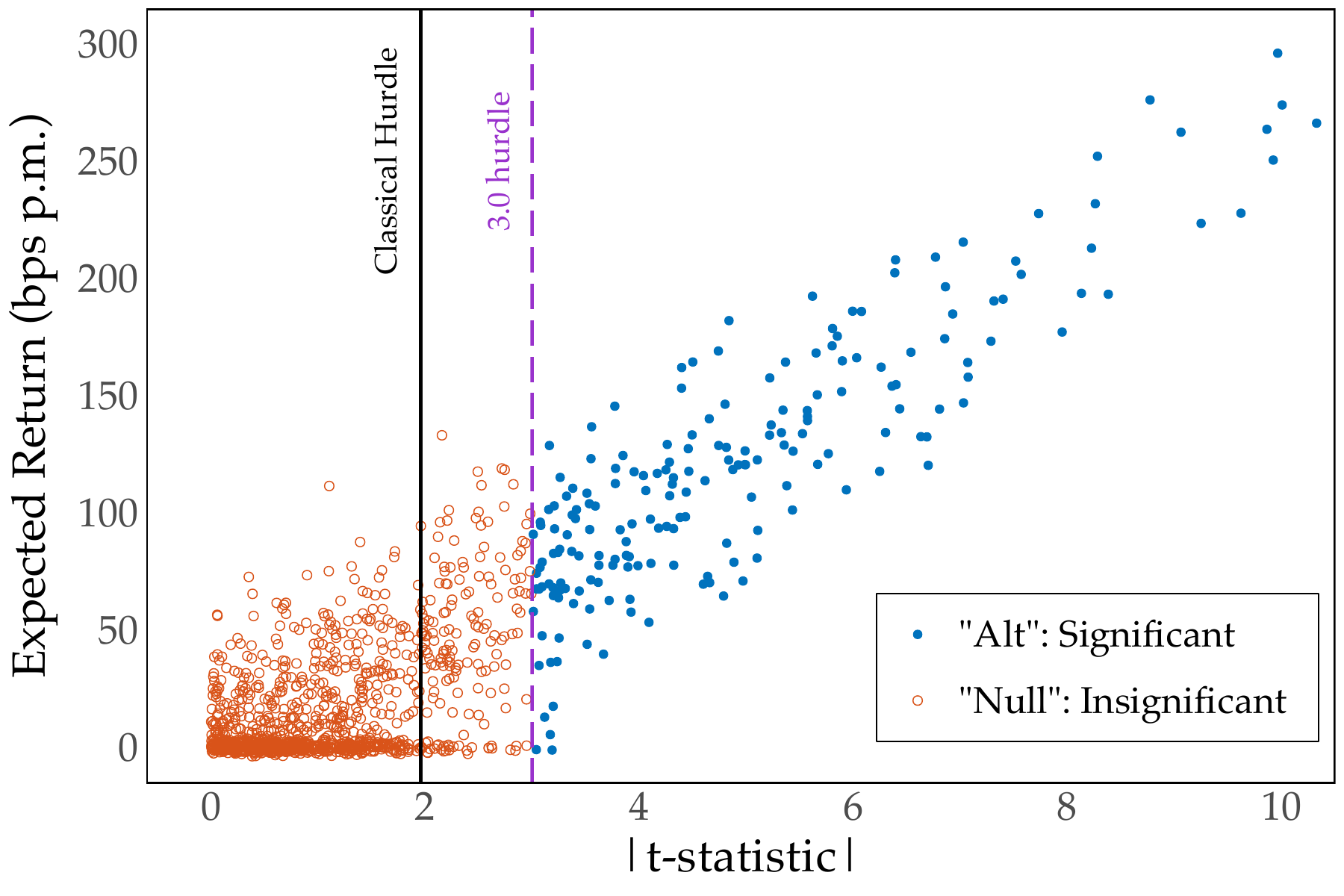
    }}
\end{figure}
\fi

Panel (a) follows the standard interpretation, labelling factors as ``Null'' (red circles) or ``Alt'' (blue dots), depending on their position on the y-axis (``Expected Return''). Significance is determined by the position on the x-axis (``$|$t-statistic$|$''). 355 factors fall to the right of the classical 2.0 hurdle. 30 of these are null, implying an FDR of only 30/355 $\approx$ 9\% using the classical hurdle for significance. This number is consistent with the Easy Bound applied to HLZ's summary stats (Table \ref{tab:ez-combined}) as well as the Visual Bound based on CLZ (Figure \ref{fig:viz}). HLZ suggest that one should aim for an FDR of 1\% (page 24), which one can obtain with a significance hurdle of $|t_i| > 3.0$ (dashed line). Nevertheless, ``null'' and ``significant'' are distinct concepts, and require separate axes for plotting.

Panel (b) follows HLZ's interpretation. Using HLZ's preferred 3.0 hurdle, factors that fall to the left of the dashed line are ``null,'' regardless of their position on the y-axis. As seen in Equations \eqref{eq:hlz-1}-\eqref{eq:hlz-3}, the y-axis is not the sample mean return, which may be due to luck, but the fundamental expected return, cured of sampling error. Thus, HLZ's interpretation leads to questionable economic conclusions, like the idea that a factor with an expected return of 100 bps per month (12\% per year) is a ``false discovery.''

Panel (b) shows that, even using HLZ's interpretation of ``false discovery,'' it is difficult to conclude that most findings are false. Of the factors that meet the classical 2.0 hurdle, 44\% fall to the left of the 3.0 hurdle. Under HLZ's interpretation, this number implies that ``many'' findings are ``false,'' or that about half are ``false.'' But it does not quite permit the conclusion that \emph{most} findings are false.

To complete the reconciliation, it helps to read HLZ's conclusion (page 37), which refers to \citet{benjamini2001control} (BY) Theorem 1.3:
\begin{quotation}
    \emph{In medical research, the recognition of the multiple testing
    problem has led to the disturbing conclusion that \textquotedblleft most
    claimed research findings are false\textquotedblright{} (\citet{ioannidis2005most}).
    Our analysis of factor discoveries leads to the same conclusion\textendash many
    of the factors discovered in the field of finance are likely false
    discoveries: of the 296 published significant factors, 158 would be
    considered false discoveries under Bonferonni [sic], 142 under
    Holm, 132 under [BY Theorem 1.3] (1\%), and 80 under [BY Theorem 1.3] (5\%).}
\end{quotation}
This quote presents four false discovery rate estimates: 53\% (158/296), 48\% (142/296), 45\% (132/296), and 27\% (80/296), and summarizes them by saying ``\emph{many}'' discoveries are likely false. Yet it also arrives at the ``same'' conclusion as \citet{ioannidis2005most}, which argues that \emph{most} claimed research findings are false. 

The final step of this argument either selects only the largest out of the four estimates (53\%) or makes an interpretive leap from ``many'' to ``most.'' Either way, HLZ's conclusion that ``most claimed research findings are likely false'' requires an additional nonstandard interpretation beyond equating ``insignificant'' and ``null.''

\subsection{Reconciliation with \citet{harvey2020false} (HL)}\label{sec:discuss-hl}

HL also examine the YZ data using \citet{storey2002direct}. But they come to a very different conclusion (page 2531):
\begin{quotation}
    \emph{Our results suggest that only about 0.1\% (or 18) of the 18,000 anomaly strategies in \citet{yan2017fundamental} are classified as ``true'' to control the FDR at 10\%.}
\end{quotation}
This 0.1\% is two orders of magnitude smaller than the 12.2\% to 49.2\% share of ``true'' (more precisely, non-null) strategies in Table \ref{tab:yz-fdr}, which also uses the YZ data and \citet{storey2002direct}.

The reconciliation once again centers on interpretations of significance. To understand the interpretations, generalize Equation \eqref{eq:viz-fdr-bound} as follows:
\begin{align}
    \text{FDR}(h) \le
    \left[
        \frac{\Pr\left(|t_{i}|>h\bigmid\nullt_i\right)}{\Pr\left(|t_{i}|>h\right)}
    \right]
    \left[
        \frac{\Pr\left(|t_i| < c \right)}{\Pr\left(|t_i| < c \bigmid\nullt_i \right)}
    \right],
    \label{eq:HL-1}
\end{align}
where $|t_i| > h$ generalizes $|t_i|>2$, $|t_i| < c$ generalizes $|t_i| < 0.5$, and $\text{FDR}(h)$ is the false discovery rate from defining significance with $|t_i| > h$. \citet{storey2002direct} uses Equation \eqref{eq:HL-1} to construct a hypothesis test: choose a value for $c$ and a desired worst-case FDR $q$. Then solve for the smallest $h$ that ensures the RHS is less than $q$. 

In the standard interpretation, strategies that satisfy $|t_i| > h$ are ``significant'' when ``controlling'' the FDR at $q$ (e.g. \citealt{storey2003statistical}). But HL follow HLZ's interpretation, and describe strategies that satisfy $|t_i| > h$ as ``true.''

Table \ref{tab:yz-fdr} Panel (b) takes a closer look, by fixing $c=1.28$ (equivalent to HLZ's  $\theta=0.8$ on page 2531) and solving for the smallest $h$ that ensures the RHS is less than 10\%. It examines all performance measures from YZ's Table 1, which includes both equal- and value-weighted specifications. The resulting hurdles vary widely, as do the shares of strategies that meet the hurdles. 

The table uses the standard interpretation, and reports that 30\% to 40\% of strategies are ``significant'' using equal-weighting, but only about 10\% are significant under most value-weighted specifications. These significance rates are smaller than the share of non-null strategies, which is roughly 50\% under equal-weighting, and ranges from 12\% to 32\% under value-weighting. 

In HLZ's interpretation, 30\% to 40\% of equal-weighted strategies are ``true,'' and about 10\% are true under most value-weighted specifications. This helps with the reconciliation, but these numbers are still two orders of magnitude larger than HL's 0.1\%. 

To reconcile further, note there is an outlier in Table \ref{tab:yz-fdr}: if one measures predictability using value-weighting \emph{and} adjusts for FF3 + Momentum exposure, only 1.0\% of strategies are ``true'' using HLZ's interpretation. This 1.0\% is almost the same magnitude as HL's 0.1\% and is consistent with HL's footnote 18, which states ``[w]e focus only on the Fama-French-Carhart four-factor model to save space.'' 

Thus, one can largely reconcile HL's claims by interpreting ``significant'' as ``true'' and selecting the smallest estimate based on the six specifications in YZ.

\subsection{Reconciliation with \citet{chordia2020anomalies}}\label{sec:discuss-cgs}

CGS report the headline FDR number in their abstract:
\begin{quotation}\noindent
    \emph{We estimate the expected proportion of false rejections that researchers
    would produce if they failed to account for multiple hypothesis testing to be about 45\%.}
\end{quotation}
This conflicts with Equation \eqref{eq:ez-dm}, which shows that this proportion is at most 25\%.

CGS use standard statistical language. Throughout their paper, a ``rejection'' is a discovery, and a false rejection is a discovery that is null (e.g. pages 2135 and 2153). So the reconciliation is not in the interpretation.

Instead, the reconciliation lies in CGS's calibration procedure. The headline number comes from a model of return signals, in which $\pi$ fraction of signals are not null (``informative,'' page 2142). They calibrate $\pi$ to their data-mined returns, arriving at $\pi = 0.06\%$ (page 2154), implying $\Pr\left(\nullt_i\right) = 99.94\%$.

$\Pr\left(\nullt_i\right) = 99.94\%$ is difficult to square with the law of total probability and figures reported in CGS. As shown in Equation \eqref{eq:viz-2}, total probability implies bounds on $\Pr\left(\nullt_i\right)$ of the form:
\begin{align}\label{eq:cgs-numbers-0}
    \Pr\left(\nullt_i\right)
    &\le \frac{\Pr\left(
        |t_i|< c
        \right)
    }{
        \Pr\left(|t_i|< c\bigmid\nullt_i\right)
    } 
\end{align}
where $c$ is a constant. CGS show in Figure 1 that at least 100,000 of 2.39 million t-stats fall in the bin containing zero. Inspecting the figure closely shows that at most 11 bins fit between zero and 2.0, and thus the width of the bin that contains zero is at least $2/12 = 0.166$. Plugging these values into Equation \eqref{eq:cgs-numbers-0} leads to
\begin{align}
    \Pr\left(\nullt_i\right)
    &\le \frac{ 
        \left(1 \times 10^5\right) / \left(2.39 \times 10^6\right)
    }{
        \Pr\left(|t_i|< (0.166/2)
        \bigmid
         \nullt_i\right)
    }
    = 63\%,
\end{align}
not far from the bounds in Table \ref{tab:yz-fdr}, and far lower than the 99.94\% calibrated by CGS. The intuition can be seen by applying Algorithm \ref{alg:viz} to Figure 1 of CGS. The largest standard normal distribution that still fits inside this histogram will fail to cover the tails, which extend well beyond -2.0 and +2.0.

So how does CGS's calibration arrive at $\pi = 0.06\%$? They calibrate $\pi$ to match the share of signals that exceed the $|t_i|>2$  threshold but fail a more stringent threshold based on \citet{romano2007control} (page 2153). This target moment contains no information about small t-stats, which, as seen in Figure \ref{fig:viz}, are highly informative about $\Pr\left(\nullt_i\right)$. As shown in Equation \eqref{eq:cgs-numbers-0}, if CGS had instead targeted the share of small t-stats, they would have arrived at a much higher $\pi$. Since CGS choose the dispersion of informative alphas outside of their calibration (page 2152), this much higher $\pi$ would then imply a much lower headline FDR.

\subsection{Reconciliation with Out-of-Sample (OOS) Tests}\label{sec:discuss-oos}

Assuming no structural changes, the mean OOS return of a false discovery is zero. Thus, if the FDR is more than $X\%$, we would expect at least $X\%$ return decay in OOS tests.

At a cursory glance, the literature on OOS tests finds mixed evidence, with decay ranging from 0\% (\citealt{jacobs2020anomalies}) to 60\% (\citealt{linnainmaa2018history}). But a closer look reveals that the decay is consistently less than 36\%, if one focuses on OOS tests that are temporally and methodologically close to the original tests. Using OOS periods that end at the publication dates, and using portfolio tests that generally follow the original papers, \citet{chen2020publication}, \citet{mclean2016does}, and \citet{jacobs2020anomalies} find OOS decay of 20\%, 26\%, and 36\%, respectively. These numbers suggest that at least 36\% of discoveries are true.

Conceptually, OOS tests that are close to the original tests are the most appropriate for validating FDR estimates. OOS samples that are far from the original samples introduce economic effects that are distinct from multiple testing. For example, improvements in information technology can reduce predictability (\citealt{chordia2014have}), while the introduction of retail traders can increase it (\citealt{goetzmann2018momentum}). Though these effects are important, they are economic rather than statistical, and should not be measured using FDR methods.

With the large datasets available from data mining, one can examine OOS periods that are even closer to the in-sample periods. CLZ find OOS decay as small as 17\% among data-mined predictors by zooming in on the 12-months after the in-sample periods. YZ and \citet{chen2024high} use data mining to produce OOS returns of roughly 50 bps per month. This mild decay and strong OOS performance are quite consistent with the $\FDRez \le 25\%$ derived in this paper.
 
\section{Conclusion and Limitations}

This paper provides simple arguments that bound the FDR under publication bias. They handle publication bias by using data-mining experiments or conservative extrapolations as worst-case scenarios. The arguments can be explained in just a few equations and provide clarity to the FDR literature. Using data from many papers, they show that the FDR cannot exceed 25\%. This small FDR  comes from a surprising fact: one out of five random accounting ratios produces statistically-significant predictability. If most discoveries were false, finding predictability would not be so easy.

Claims that the FDR exceeds 25\% are driven by interpreting ``insignificant factors'' as ``false discoveries'' and selecting the extremes among multiple estimates. Neglecting to fit moments that are highly informative about the FDR also results in overestimates.

As highlighted by the title of this paper, my analysis is limited to statistics. A ``true discovery'' is a statistically significant predictor that has a non-zero alpha. Many economic issues are buried in this statement, like trading costs (\citet{novy2016taxonomy}), information costs (\citet{chordia2014have}), and arbitrage effects (\citet{mclean2016does}). \citet{chen2023zeroing} find that, net of these effects, most findings in cross-sectional predictability produce measly profits---even if they are true, statistically speaking.

\pagebreak{}

\ifshowreferences
    \pdfbookmark{References}{bib}
    \printbibliography
\fi


\end{document}